\documentclass{IEEEoj}

\usepackage{cite}
\usepackage{amsmath,amssymb,amsfonts}
\usepackage{algorithmic,mathrsfs}
\usepackage{graphicx,subfigure}
\usepackage{textcomp,bm,tabularx}
\usepackage{balance}
\usepackage{soul,comment,tikz}
\usepackage{color,colortbl,setspace}
\usepackage{tabularx,booktabs}
\usepackage{steinmetz,relsize}
\usepackage{xcolor}
\usepackage[hyphens]{url}

\definecolor{Gray}{gray}{0.92}
\definecolor{mygreen}{RGB}{0,162,15}

\def\BibTeX{{\rm B\kern-.05em{\sc i\kern-.025em b}\kern-.08em
T\kern-.1667em\lower.7ex\hbox{E}\kern-.125emX}}
\AtBeginDocument{\definecolor{ojcolor}{cmyk}{0.93,0.59,0.15,0.02}}
\def\OJlogo{\vspace{-14pt}\includegraphics[height=20pt]{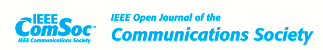}}

\let\labelindent\relax
\usepackage{enumitem}

\usepackage{scalerel}
\usepackage{tikz}
\usetikzlibrary{svg.path}

\definecolor{orcidlogocol}{HTML}{A6CE39}
\tikzset{
  orcidlogo/.pic={
    \fill[orcidlogocol] svg{M256,128c0,70.7-57.3,128-128,128C57.3,256,0,198.7,0,128C0,57.3,57.3,0,128,0C198.7,0,256,57.3,256,128z};
    \fill[white] svg{M86.3,186.2H70.9V79.1h15.4v48.4V186.2z}
                 svg{M108.9,79.1h41.6c39.6,0,57,28.3,57,53.6c0,27.5-21.5,53.6-56.8,53.6h-41.8V79.1z M124.3,172.4h24.5c34.9,0,42.9-26.5,42.9-39.7c0-21.5-13.7-39.7-43.7-39.7h-23.7V172.4z}
                 svg{M88.7,56.8c0,5.5-4.5,10.1-10.1,10.1c-5.6,0-10.1-4.6-10.1-10.1c0-5.6,4.5-10.1,10.1-10.1C84.2,46.7,88.7,51.3,88.7,56.8z};
  }
}

\newcommand\orcidicon[1]{\href{https://orcid.org/0000-0002-9393-7104}{\mbox{\scalerel*{
\begin{tikzpicture}[yscale=-1,transform shape]
\pic{orcidlogo};
\end{tikzpicture}
}{|}}}}

\newcommand*\circled[1]{\tikz[baseline=(char.base)]{\node[shape=circle,draw,inner sep=1pt] (char) {#1};}}

\usepackage[hidelinks]{hyperref} 

\begin{document}

\receiveddate{01 June, 2023}
\reviseddate{XX July, 2023}
\accepteddate{XX Month, XXXX}
\publisheddate{XX Month, XXXX}
\currentdate{XX Month, XXXX}
\doiinfo{OJCOMS.2022.1234567}

\title{Assessing Adversarial Replay and Deep Learning-Driven Attacks on Specific Emitter Identification-based Security Approaches}

\author{Joshua H. Tyler\authorrefmark{1}, STUDENT MEMBER, IEEE, Mohamed K. M. Fadul\authorrefmark{1}, Matthew R. Hilling\authorrefmark{1}, Donald R. Reising\authorrefmark{1}\orcidicon, SENIOR MEMBER, IEEE, AND T. Daniel Loveless\authorrefmark{1}, SENIOR MEMBER, IEEE}
\affil{Electrical Engineering Department, The University of Tennessee at Chattanooga, Chattanooga, 
TN 37403 USA}
\corresp{CORRESPONDING AUTHOR: Donald R. Reising (e-mail: donald-reising@utc.edu).}
\markboth{Assessing Adversarial Replay and Deep Learning-Driven Attacks on Specific Emitter Identification-based Security Approaches}{Joshua H. Tyler \textit{et al.}}

\begin{abstract}
Specific Emitter Identification (SEI) detects, characterizes, and identifies emitters by exploiting distinct, inherent, and unintentional features in their transmitted signals. Since its introduction a significant amount of work has been conducted; however, most assume the emitters are passive and that their identifying signal features are immutable and challenging to mimic. Suggesting the emitters are reluctant and incapable of developing and implementing effective SEI countermeasures; however, Deep Learning (DL) has been shown capable of learning emitter-specific features directly from their raw in-phase and quadrature signal samples and Software-Defined Radios (SDRs) are capable of manipulating them. Based on these capabilities, it is fair to question the ease at which an emitter can effectively mimic the SEI features of another or manipulate its own to hinder or defeat SEI. This work considers SEI mimicry using three signal features mimicking countermeasures; “off-the-self” DL; two SDRs of different sizes, weights, power, and cost (SWaP-C); handcrafted and DL-based SEI processes, and a ``coffee shop'' deployment. Our results show “off-the-shelf” DL algorithms and SDR enables SEI mimicry; however, adversary success is hindered by: (i) the use of decoy emitter {preambles}, (ii) the use of a denoising autoencoder, and (iii) SDR SWaP-C constraints.
\end{abstract}



\maketitle

\section{Introduction}
\label{sec:introduction}
\IEEEPARstart{S}{pecific Emitter Identification (SEI)} was introduced almost thirty years ago to provide electronic warfare systems the capacity to detect, characterize, and identify radars by exploiting distinct, inherent, and unintentional features present in their transmitted signals~\cite{langley1993specific}. Over that time there has been a high volume of SEI research conducted. More recently SEI has seen a surge of interest due to Internet of Things (IoT) security concerns and the emergence of Deep Learning (DL). SEI is an advantageous IoT security approach because it passively exploits features that are unintentionally imparted to the signal--by the emitter's Radio Frequency (RF) front-end circuitry--during its formation and transmission, thus eliminating the need to modify or add IoT device functionality. Moreover, SEI's passive nature makes it a viable security mechanism for resource-constrained IoT devices (e.g., limited memory, computation, or battery). Additionally, DL-based SEI has shown the capability to learn emitter-specific features directly from the raw In-phase and Quadrature (IQ) signal samples, thus eliminating expert-reliant feature engineering~\cite{restuccia2019deepradioid}.\\
\indent Most SEI research assumes the identified emitters are passive and their identifying features immutable and difficult to mimic, which suggests they are reluctant and incapable of developing and implementing effective SEI countermeasures~\cite{xu2015device,han2020radio}. However, when considering DL's ability to learn emitter-specific features directly from the raw IQ signal samples and coupling it with readily available ``off-the-shelf'' DL architectures and the flexibility of Software-Defined Radio (SDR) one must reconsider the degree to which these assumptions hold true~\cite{tensorflow2015-whitepaper,sadiku2004software}. Our objective is to assess an emitter's ability to easily and effectively mimic another emitter's distinct, inherent, and unintentional signal features by employing DL to learn them and SDR to manipulate or mask its own for the purpose of hindering or defeating SEI. Prior works have investigated SEI mimicry~\cite{danev2010attacks,restuccia2020hacking,shi2020generative,karunaratne2021penetrating}. Our work differs from these works and extends the community's understanding of SEI mimicry in the following ways.
\begin{itemize}
\item{Performs SEI mimicry using ``off-the-shelf'' DL algorithms while prior efforts use algorithms tailored specifically with SEI mimicry in mind. Our aim is to determine the ease at which SEI mimicry can be implemented.}
\item{Conducts SEI mimicry using two SDRs that differ in Size, Weight, Power, and Cost (a.k.a., SWaP-C) to determine how SDR capabilities impact SEI mimicry performance.}
\item{Targets mimicry of non-SDR emitters. Specifically, the emitters being mimicked are eight TP-Link Archer T3U USB 802.11a Wireless-Fidelity (Wi-Fi) compliant emitters, which better represent Internet of Things (IoT) deployed and user devices.}
\item{Uses an adversary that conducts SEI mimicry using {preamble} replay, an AutoEncoder (AE), or convolutional Generative Adversarial Network (GAN) to mimic the SEI features present within the signals of another emitter.}
\item{Assesses SEI mimicry effectiveness in hindering or defeating handcrafted (a.k.a., feature-engineered) and DL-based SEI processes to enable comparative assessment that is not seen in prior works.}
\item{Considers the presence of ``decoy'' emitter signals to aid the SEI process in discerning the adversary from the authorized emitter being mimicked. A decoy emitter is an SDR of the same manufacturer and model as that used by the adversary but has a different serial number.}
\item{Assesses the effectiveness of GAN-based SEI mimicry within a realistic, in-situ ``coffee shop'' scenario, which is unseen in previous works.}
\item{Performs SEI and mimicry countermeasure assessments using handcrafted- and DL-based processes to facilitate comparative assessment that is unseen in published literature.}
\end{itemize} 

\noindent Our results show ``off-the-shelf'' DL algorithms and SDR enables SEI mimicry; however, adversary success is hindered by: (i) the use of decoy emitter signals, (ii) integration of a Denoising AE (DAE), and (iii) SDR SWaP-C constraints.\\
\indent The article is organized as follows. Section~\ref{sec:threat_model} defines the threat model while Section~\ref{sec:background} describes the signal of interest, and RF-Distinct, Native Attributes (RF-DNA) Fingerprinting, and the machine learning architectures used. Section~\ref{sec:methods} describes the signal collection, detection, and preprocessing, DL configurations, as well as the ``coffee shop'' scenario, Section~\ref{sec:results} presents the results and analysis while the conclusion is presented in Section~\ref{sec:conlusion}.
\section{THREAT MODEL%
\label{sec:threat_model}}
The threat model used herein is an extension of the threat model presented in our prior work~\cite{reising2020radio}. In that work, the adversary (a.k.a., Eve) attempts to gain unauthorized network access by assuming the digital identity (e.g., MAC address) of an authorized (a.k.a., Alice) device such that it is granted unauthorized network access by being incorrectly authenticated by the network monitor (a.k.a., Bob). It is assumed that Eve conducts the attack using simple software tools, is not an authorized device, and does not inherently have access to the network or its associated devices. Finally, it is assumed that network communication links and hardware are not originally compromised.\\
\indent Eve's capabilities are extended to include the use of ``off-the-shelf'' DL and Commercial-Off-The-Shelf (COTS) SDRs to develop and implement a mimicry countermeasure for the purpose of hindering or defeating SEI. Eve is implemented using two SDR platforms to determine SWaP-C impacts upon SEI mimicry success. The two SDR platforms are Ettus Research's Universal Software Radio Peripheral (USRP) B210 ($\sim$\$4,000 per unit) and Great Scott Gadget's HackRF One ($\sim$\$470 per unit)~\cite{usrp_b210,hackrf_one}. 
Eve's use of the full-duplex B210 is advantageous because it can receive its own transmitted signals while the half-duplex HackRF cannot. Thus, the HackRF requires Eve to employ a second one to receive the signals transmitted by the first, which increases Eve's complexity. Eve employs one of the following SEI countermeasures. 
\begin{itemize}
\item{\textit{\underline{Replay SEI Mimicry}}: Eve collects, saves, and re-transmits an authorized emitter's signals. Eve can modify the replayed signals by adjusting its transmit power and CFO behavior, but neither is not assessed here.
}
\item{\textit{\underline{AE-based SEI Mimicry:}} Eve collects signals transmitted by Alice and trains an AE such that when Eve inputs one of its own signals into the trained AE it ``recolors'' the signal's features to match those present in Alice's signals. The AE has a hidden layer of size one hundred and a targeted Mean Squared Error (MSE) of $1$$\times$$10^{-6}$~\cite{tyler2021simplified}.
}
\item{\textit{\underline{GAN-based SEI Mimicry:}} Eve collects its own and Alice's signals. The GAN is trained by assigning Alice's collected preambles to Class~{\#1} of the discriminator $D$ and its own preambles to Class~{\#2}. During training, Class~{\#2} is input to the generator $G$, and its weights are adjusted to learn the mapping needed to modify the SEI features--present in Eve's signals--to match those of Alice (a.k.a., Class~{\#1}) at the $G$'s output. The success of this mapping is determined by how often the $D$ assigns the $G$'s output to Class~{\#1}. The $D$ is a basic, three-layer CNN and trained to discriminate Class~{\#1}'s signals from those of Class~{\#2} modified by the $G$. The $G$ and $D$ are iteratively updated until the $D$ can no longer discern a Class~{\#1} signal from those output by the $G$. Ideally, the $G$ learns to remove Class~{\#2}'s inherent SEI features and insert those of Class~{\#1}.}
\end{itemize}
The AE and GAN are trained using unit energy signals to ensure the correct feature distributions are learned by the associated DL network without signal energy biasing the process. Only the signals' raw IQ representations are used to train the AE and GAN. The AE and GAN-based SEI countermeasures are implemented by passing the signal's IQ samples through the trained AE and $G$ and their outputs are stored for later transmission by Eve.

Eve employs each SEI countermeasure in conjunction with a falsified digital identity to increase its chance of being incorrectly authenticated by Bob. Bob verifies the identity of the to-be-authenticated device using the passed digital identity and the associated received signals' SEI features.
\section{BACKGROUND%
\label{sec:background}}
\subsection{Signal of Interest%
\label{sec:sig_of_interest}}
\begin{figure}[!b]
\centering
\includegraphics[width=0.95\columnwidth]{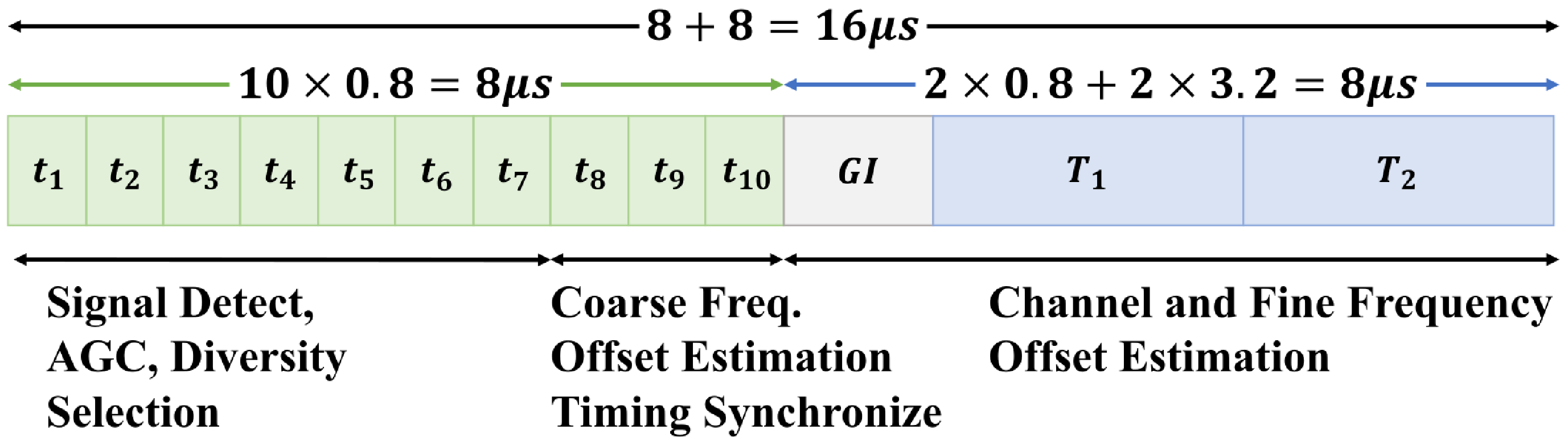}
\caption{The structure of the IEEE 802.11a Wi-Fi preamble that occupies the first 16~{$\mu$s} of every transmitted signal~\cite{ieee2007ieee}.}
\label{fig:preamble}
\end{figure}
This work uses IEEE 802.11a Wi-Fi-compliant signals. IEEE 802.11a Wi-Fi is an Orthogonal Frequency Division Multiplexed (OFDM) digital encoding scheme. IEEE 802.11a Wi-Fi is used in this work for the following reasons: (i) the first 16~{$\mu$s} of every IEEE 802.11a Wi-Fi frame consists of a preamble. The preamble enables synchronization, carrier frequency offset correction, phase offset correction, and channel equalization at the receiver; (ii) the IEEE 802.11a Wi-Fi preamble consists of a fixed and known sequence of symbols, thus making it ideal for the extraction of steady-state signal features by an SEI process. An illustration of the IEEE 802.11a Wi-Fi preamble structure is shown in Figure~\ref{fig:preamble}; (iii) the processes and techniques presented herein can be extended to other preamble-based standards such as ZigBee and Bluetooth; (iv) the number of publications demonstrating successful SEI exploitation of IEEE 802.11a Wi-Fi preamble extracted features~\cite{jeffrey2013802,suski2008radio,liu2008specific,takahashi2010ieee,liu2011nonlinearity,reising2015authorized,wheeler2017assessment,fadul2019rf,fadul2021nelder,riyaz2018deep,restuccia2019deepradioid,tyler2021simplified,tyler2022analysis}. These works have achieved serial number discrimination (i.e., same manufacturer and model number), which is the most challenging SEI case; (v) the use of OFDM or a variation of OFDM (e.g., Orthogonal Frequency Division Multiple Access) in many digital communications schemes. These schemes include IEEE 802.11ac, IEEE 802.11ad, IEEE 802.11ax, and Long-Term Evolution (LTE).
\subsection{RF-DNA Fingerprinting%
\label{sec:rf_dna_fingerprinting}}
\subsubsection{Handcrafted RF-DNA Fingerprinting%
\label{sec:hand_rf_dna}}
\begin{figure}[!t]
\centering
\includegraphics[width=\columnwidth]{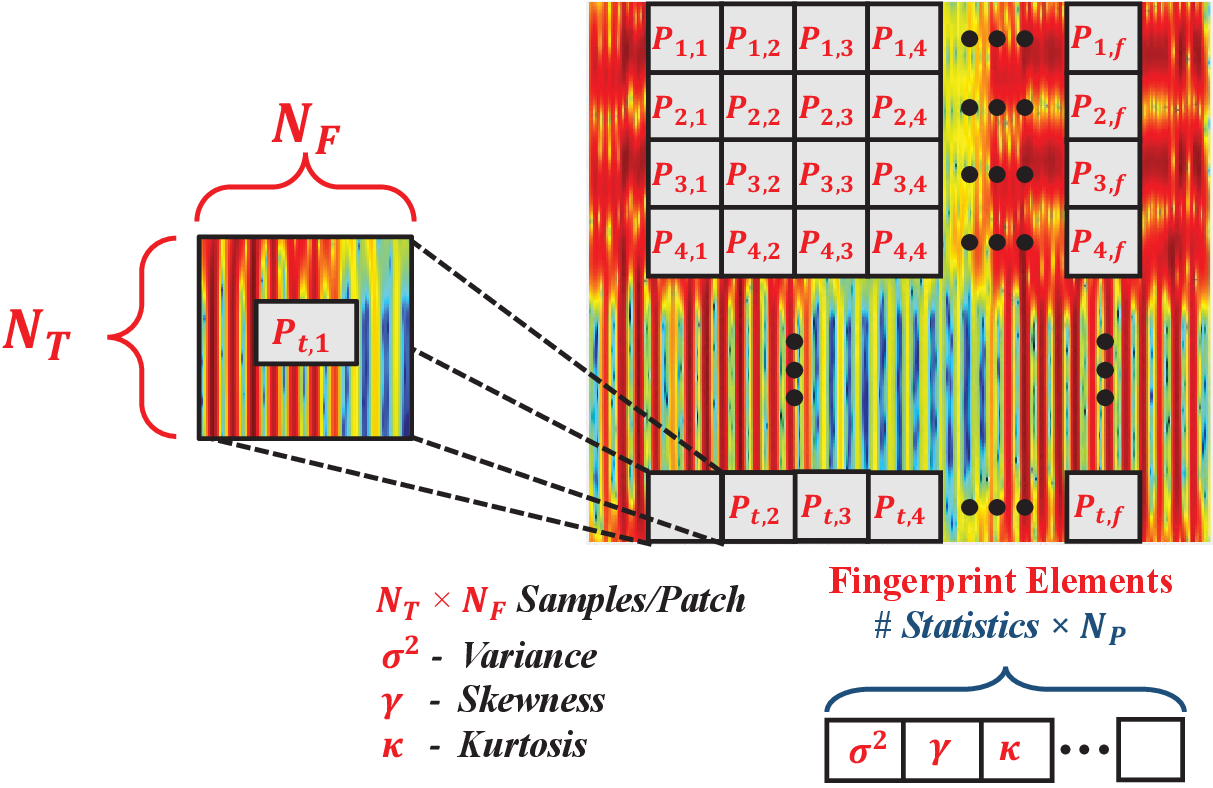}
\caption{Gabor-based RF-DNA fingerprint generation is detailed in~\cite{reising2015authorized} and adopted for time-frequency-based SEI.}
\vspace{-10mm}
\label{fig:gt_based_rf_dna}
\end{figure}

Handcrafted RF-DNA fingerprints are extracted from the Wi-Fi preamble's two-dimensional (2D) Time-Frequency (TF) representation, which is generated using the Discrete Gabor Transform (DGT)~\cite{bastiaans1996discrete}. The TF representation is the squared magnitude of the complex-valued Gabor coefficients and normalizing the resulting surface such that all of its values are in the range of zero to one. This surface is subdivided into $N_{P}$ ``patches'' and the patch size is chosen to ensure that each patch contains at least fifteen elements. Each patch is converted into a one-dimensional (1D) vector and variance $(\sigma^{2})$, skewness $(\gamma)$, and kurtosis $(\kappa)$ calculated. This process is repeated for each patch over the entire TF surface and the statistics are appended to those of the previous patch. Figure~\ref{fig:gt_based_rf_dna} shows the subdivision of the normalized, magnitude-squared Gabor-based surface into $N_{P}$ patches, statistics calculated, and appended to form a handcrafted RF-DNA fingerprint. The three statistics are calculated over the entire TF surface and added to the end of the last patch's statistics. The reader is directed to our work in~\cite{fadul2023analysis} for a detailed explanation and the specific values used to compute the DGT and handcrafted RF-DNA fingerprints.
\subsubsection{Deep Learning-based RF-DNA Fingerprinting%
\label{sec:dl_rf_finger}}
DL-based RF-DNA fingerprinting uses the preamble's time, frequency, or Gabor-based representation. The time and frequency representation is from our work in~\cite{tyler2021simplified,tyler2022analysis}. For the time representation, the CNN learns RF-DNA fingerprints from the preamble's augmented IQ samples,
\begin{equation}
\bm{P}_{e,k} = \begin{bmatrix}
i_{e,k}(1,1) & i_{e,k}(1,2) & \cdots & i_{e,k}^{i}(1,N_{s}) \\
q_{e,k}(2,1) & q_{e,k}(2,2) & \cdots & q_{e,k}(2,N_{s}) \\
\lambda_{e,k}(3,1) &  \lambda_{e,k}(3,2) & \cdots &  \lambda_{e,k}(3,N_{s}) \\
\theta_{e,k}(4,1) & \theta_{e,k}(4,2) & \cdots &  \theta_{e,k}(4,N_{s})
\end{bmatrix},
\label{eqn:preamble_time_rep}
\end{equation}
where $k$ is the index assigned to the emitter's $k^{\text{th}}$ received preamble $\bm{r}_{e}$, $N_{s}$ is the number of discrete-time samples comprising the emitter's $k^{\text{th}}$ preamble, $\bm{i}$ are the in-phase components, $\bm{q}$ are the quadrature components, $\bm{\lambda}$ and $\bm{\theta}$ are the instantaneous magnitude and phase of $\bm{r}_{e}$. The frequency representation is generated using $\bm{r}_{e}$'s discrete Fourier transform's complex-valued coefficients and is expressed as,
\begin{equation}
\resizebox{.9\hsize}{!}{$
\bm{\mathscr{P}}_{e,k} = \begin{bmatrix}
I_{e,k}(1,1) & I_{e,k}(1,2) & \cdots & I_{e,k}(1,N_{s}) \\
Q_{e,k}(2,1) & Q_{e,k}(2,2) & \cdots & Q_{e,k}(2,N_{s}) \\
\Lambda_{e,k}(3,1) &  \Lambda_{e,k}(3,2) & \cdots &  \Lambda_{e,k}(3,N_{s}) \\
\Theta_{e,k}(4,1) & \Theta_{e,k}(4,2) & \cdots &  \Theta_{e,k}(4,N_{s})
\end{bmatrix}$},
\label{eqn:preamble_freq_rep}
\end{equation}
where $\bm{I}$ and $\bm{Q}$ are the real and imaginary components (a.k.a., the rectangular representation) of the complex-valued Fourier coefficients, respectively. $\bm{\Lambda}$ and $\bm{\Theta}$ are the magnitude and phase responses (a.k.a., the polar representation) of the complex-valued Fourier coefficients, respectively.\\
\indent The Gabor-based representation is the TF surface--in Section~\ref{sec:hand_rf_dna} before patch subdivision--converted into a Red, Green, and Blue (RGB) image that captures time, frequency, and color-based intensity in a three-dimensional tensor.
\subsection{Machine Learning Architectures%
\label{sec:ml_arch}}

\subsubsection{Multiple Discriminant Analysis/Maximum Likelihood%
\label{sec:dl_arch_mdaml}}
Multiple Discriminant Analysis (MDA) reduces feature dimensionality while improving class separability. MDA extends Fisher's linear discriminant analysis from a two-class case to the $C$-class case, where $C$ is the number of known emitters. MDA is a linear operation that projects the handcrafted RF-DNA fingerprints into a ($C-1$)-dimensional subspace without reducing class separability by maximizing inter-class distances while minimizing intra-class spread~\cite{theodoridis2006pattern}.

Maximum Likelihood (ML) classification assigns each MDA-projected RF-DNA fingerprint to one of the $C$ possible classes using Bayesian Decision Theory. A classification decision is made by computing the likelihood value between the multivariate Gaussian distribution fit to each class and the \textit{unknown} MDA-projected RF-DNA fingerprint. The \textit{unknown} MDA-projected RF-DNA fingerprint is assigned to the class whose distribution results in the largest likelihood value. The classification error probability is minimized by assuming uniform costs and equal priors~\cite{theodoridis2006pattern}. 
\subsubsection{Convolutional Neural Network%
\label{sec:dl_arch_cnn}}
A Convolutional Neural Network (CNN) is designed for the classification of image data, thus its first layer extracts class-specific (a.k.a., emitter-specific) features by convolving 2D filters across an input image. This feature extraction layer is referred to as a convolutional layer and the learning process optimizes its filters. The convolutional layer's output is reduced by maximum pooling (a.k.a., MaxPooling), which saves the highest value enclosed by a window that is slid across the output. The resulting feature map is smaller but still preserves most of the important data. MaxPooling is followed by a {Rectified Linear Unit (ReLU)} layer that zeros out any negative feature map values~\cite{patterson2017deep}. The use of a CNN is due to it being the most common DL architecture used in SEI as well as our own familiarity with it~\cite{fadul2021identification,tyler2021simplified,fadul2021adversarial,tyler2022analysis,tyler2022assessing}
\subsubsection{Autoencoder%
\label{sec:dl_arch_ae}}
An AE is a deep generative model that attempts to reconstruct the input data at its output layer by learning a good representation and useful properties of the data and can be described as consisting of an encoder and decoder. The encoder is responsible for generating a representation (a.k.a., a code) from the input data while the decoder regenerates the input data from the code~\cite{goodfellow2016deep}. In this work, an AE is trained by the adversary to learn the SEI feature distribution associated with Alice's signals. Once trained, the AE is used by Eve to modify the SEI features present within its own signals to match those of Alice.
\subsubsection{Generative Adversarial Network%
\label{sec:dl_arch_gan}}
A GAN learns deep representations through semi-supervised or unsupervised learning with applications to image editing, style transfer, image synthesis, and classification\cite{goodfellow2020generative,creswell2018generative}. A GAN employs adversarial training to estimate a generative model by simultaneously training two networks: (i) a Generator $G$ that captures the input data distribution, and (ii) a Discriminator $D$ that focuses on deciding if its input originated from the training data set or was created by the $G$. Thus, a GAN can be viewed as a minimax two-player game where $G$ attempts to increase the probability that $D$ makes the wrong decision. A unique solution to this minimax problem exists when: (i) the $G$ captures the training data distribution, and (ii) the probability that the $D$ makes the correct decision is one-half everywhere \cite{goodfellow2020generative}. As with the AE, Eve trains the GAN such that the $G$ learns the SEI feature distribution of Alice's signals. Unlike the AE, the feedback relationship that exists between the $G$ and $D$ networks allows the $G$ to accept Eve's signals as inputs. After training, the $G$ is disconnected from the $D$ and integrated into Eve's architecture to modify its signal's SEI features to mimic those of Alice.
\section{METHODOLOGY%
\label{sec:methods}}
\subsection{Signal Collection, Detection, and Preprocessing}
\label{sec:signal_collect}
This section describes the SEI process' and Eve's signal collection and preprocessing procedures with the exception of those used in the ``Coffee Shop'' Scenario (see Section~\ref{sec:coffee_shop_methods}). 
\subsubsection{The SEI Process}
\label{sec:signal_collect__sei}
All signals are transmitted at 5.805~{GHz} and collected using a Tektronix Real-time Spectrum Analyzer (RSA) 5126B with an ultra wide-band antenna. The Device Under Test (DUT) (a.k.a., each of the TP-Links) transmits a 2~{GB} file to an SFTP server. The signals are sampled at 200~{MHz} and the distance between the DUT and RSA antennas is set to achieve a Signal-to-Noise Ratio (SNR) of 30~{dB} or 9~{dB}, which corresponds to three or eight feet, respectively. Each collection is filtered using a fourth-order elliptical filter with a passband ripple of 0.5~{dB}, a stopband attenuation of 20~{dB}, and a cutoff frequency of 8.865~{MHz}. After filtering, an empirically selected amplitude threshold is used to detect and remove individual Wi-Fi frames from each collection record. Each frame's preamble is detected by calculating the cross-correlation between the detected frame's magnitude and the magnitude of an ideal 802.11a preamble. The cross-correlation's maximum corresponds to the midpoint of the detected frame's preamble. Following detection and extraction each preamble undergoes Carrier Frequency Offset (CFO) correction in accordance with~\cite{schmidl1997robust}, downsampling to 20~{MHz}, and energy normalization~\cite{tyler2022analysis}. Each emitter is represented by one thousand preambles.
\subsubsection{The Adversary}
\label{sec:signal_collect__eve}
Eve continuously observes the 5.805~{GHz} channel and collects all signals using either a USRP B210 or HackRF SDR at a sampling rate of 40~{MHz} or 20~{MHz}, respectively. Individual preambles are detected and extracted from Eve's collection records using the approach described in Section~\ref{sec:signal_collect__sei}. A total of one thousand preambles are collected and used to implement one of the three SEI mimicry attacks.
\subsection{{Deep Learning Architecture Configurations}}
\subsubsection{{Mimicry Networks}}
\begin{figure}[!b]
\centering
\includegraphics[width=0.7\columnwidth]{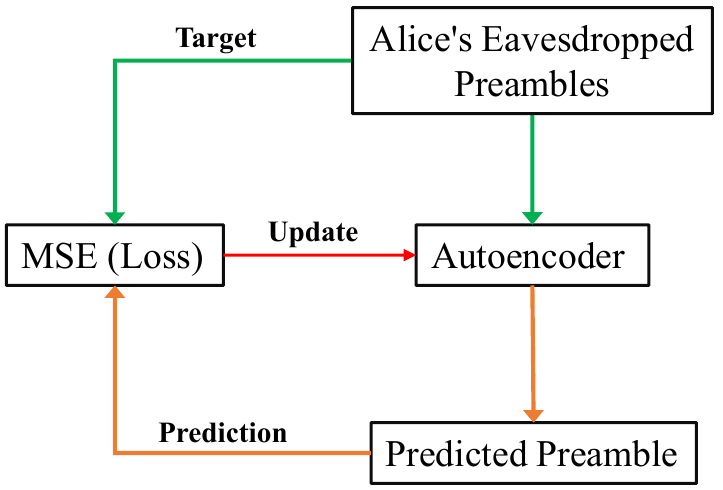}
\caption{{Flow diagram illustrating how Eve trains its AE-based mimicry network.}}
\label{fig:ae_flow_diagram}
\end{figure}

{\underline{\emph{AE Mimicry:}} Eve uses the eavesdropped preambles from Alice as the input and target for a Multi-Layer Perceptron (MLP)-based AE whose parameters are given in Table~\ref{tab:ae_arch}. The AE utilizes the real, imaginary, magnitude, and phase components of the time-domain preamble as the input and target output. Eve's AE consists of weights such that the hidden state is a vector of length $32$, and the output is a vector of length $1,280$. Each layer is activated using the sigmoid function. The AE is trained until either (i) a target MSE of $10^{-8}$ has been reached or (ii) one hundred thousand epochs have been completed. The training process for the AE-based mimicry attack is illustrated in Figure~\ref{fig:ae_flow_diagram}.}
\begin{table}[!t]
\centering
\caption{{Mimicry autoencoder architecture trained and employed by Eve.}}
\begin{tabular}{ccccc}
\toprule
Layer~{\#} & Layer & Parameters & Activation\\
\hline
\rowcolor{Gray}
1 & Dense & $128$ & Sigmoid\\
2 & Dense & $1280$ & Sigmoid\\
\bottomrule
\end{tabular}
\label{tab:ae_arch}
\end{table}
\begin{figure}[!b]
\centering
\includegraphics[width=\columnwidth]{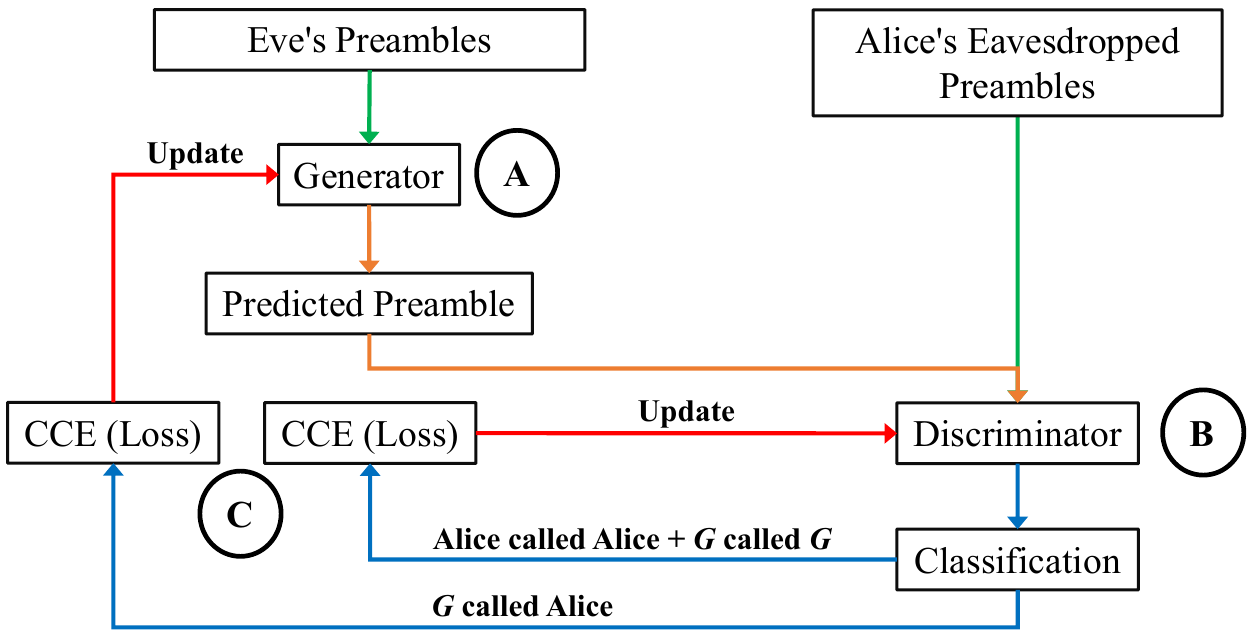}
\caption{{Flow diagram illustrating how Eve trains its GAN-based mimicry network.}}
\label{fig:gan_flow_diagram}
\end{figure}
{
\noindent \underline{\emph{GAN Mimicry:}} Eve passes her own, self-monitored preambles through a Convolutional GENerator (C-GEN) that is marked by \circled{A} in Figure~\ref{fig:gan_flow_diagram}. The C-GEN, whose parameters are provided in Table~\ref{tab:gen_arch} is comprised of (i) a two-Dimensional CONVolutional (2D-CONV) layer with sixty-four $[3\times3]$ filters, (ii) a 2D-CONV layer with one hundred and twenty-eight $[2\times4]$ filters, (iii) a MaxPooling layer with a $[2\times2]$ window, (iv) a 2D-CONV layer with sixty-four $[1\times3]$ filters, (v) a convolutional transpose layer with sixty-four $[3\times3]$ filters, (vi) a 2D upsampling layer of size $[1\times2]$, and (vii) a convolutional transpose layer with one $[1\times3]$ filters. The output of the C-GEN is a $[4\times320]$ tensor. The output of the C-GEN is passed to the Convolutional DIScriminator (C-DIS)--whose parameters are detailed in Table~\ref{tab:dis_arch}--along with the eavesdropped preambles. This step is marked in Figure~\ref{fig:gan_flow_diagram} by the \circled{B}. The C-DIS consists of (i) a 2D-CONV layer with sixty-four $[3\times3]$ filters, (ii) a 2D-CONV layer with one hundred and twenty-eight $[2\times3]$ filters, (iii) a MaxPooling layer with a $[2\times2]$ window, (iv) a 2D-CONV layer with thirty-two $[2\times3]$ filters, (v) a 2D-CONV layer with sixteen $[2\times5]$ filters, (vi) a flattening layer, (vii) a dense (a.k.a., a fully connected layer) with one-hundred and twenty-eight cells, and (viii) a dense layer with one cell. All C-GEN and C-DIS layers are activated using ReLU except for the two dense layers in the C-DIS. The first C-DIS dense layer has no activation and the second dense layer utilizes sigmoid activation. The loss functions are calculated for each network (see \circled{C} in Figure~\ref{fig:gan_flow_diagram}). The C-DIS aims to label Alice as one and the C-GEN output as zero. The C-GEN's loss function is determined by calculating the Categorical Cross Entropy (CCE) when the C-DIS classifies the C-GEN output compared to a vector of ones (i.e., the C-DIS identifies Eve as Alice). The C-DIS' loss function is determined by calculating the CCE when the C-DIS predicts Alice's preambles compared to a vector of ones (i.e., the C-DIS identifies Alice's preambles as originating from Alice), calculating the CCE when the C-DIS predicts the output of the C-GEN compared to a vector of zeros (i.e., the C-DIS identifies the C-GEN's output as coming from the C-GEN), and adding the two values. The C-GEN and C-DIS are trained on the real, imaginary, magnitude, and phase representation of the time-domain preambles. The GAN is trained until ten epochs have been reached with a minibatch size of 250.}
\begin{table}[!t]
\centering
\caption{{Specifications for the GAN architecture used by Eve to mimic the SEI features present in Alice's preambles.}} \label{tab:eve_gan_arch}
\begin{subfigure}[{Network parameters for the Generator, $G$.}]{\label{tab:gen_arch}
\includegraphics[width=\columnwidth]{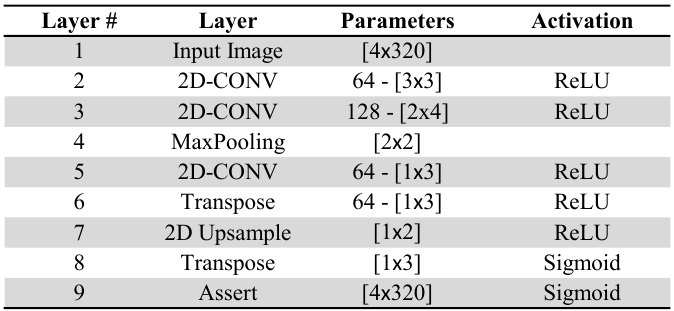}}
\end{subfigure}
\medskip
\begin{subfigure}[{Network parameters for the Discriminator, $D$.}]{\label{tab:dis_arch}
\includegraphics[width=\columnwidth]{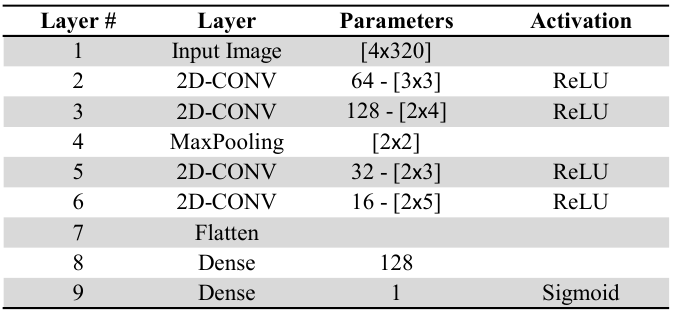}}
\end{subfigure}
\vspace{-2mm}
\end{table}
\subsubsection{SEI Network}
\label{sec:dl_configs}
\underline{\emph{Time and Frequency:}} The time and frequency CNN--whose specifications are given in Table~\ref{tab:sei_time_freq_arch}--consists of three, consecutive 2D-CONV layers. The network input is a $[4 \times 320]$ tensor. All 2D-CONV layers use a $[4 \times 4]$ filter and the number of filters per layer is $[40,80,160]$. The first 2D-CONV layer is normalized using a ReLU layer, and then a MaxPooling with a patch size of four and a stride of four. The second 2D-CONV layer is normalized using a ReLU layer, and then a MaxPooling with a patch size of one and a stride of four. All networks are trained using one hundred epochs with an $8,000$-sized minibatch.

\noindent\underline{\emph{Gabor:}} For DL-based SEI using the Gabor representation, the CNN is comprised of three layers and accepts a $[320 \times 320 \times 3]$ input tensor, Table~\ref{tab:sei_gabor_arch}. The input's third dimension captures the RGB colors of the Gabor-derived image. All 2D-CONV layers have a filter size of $[32 \times 32]$. The first 2D-CONV layer consists of eight filters while the second and third consist of sixteen and thirty-two filters, respectively. Each layer uses MaxPooling normalization with a $[32 \times 32]$ patch size, a stride of thirty-two, and ReLU activation. Each network is given thirty training epochs and an $8,000$-sized minibatch.\\
\indent Both SEI networks' classification layer is a dense layer with eight cells for no decoy, and nine cells when a decoy emitter is present within the training set. Both dense layers are activated using softmax. The networks both use CCE as their loss function. All network parameters are optimized using the~{ADAptive Momentum (ADAM)} algorithm with a learning rate of $0.01$, a gradient threshold of $1$, and an L2 regularization factor of $10^{-4}$. Networks are trained using MATLAB\textsuperscript{\textregistered}'s DL Toolbox across four compute nodes, each with two AMD Epyc 7662 64C/128T CPUs and an Nvidia A100 80GB PCI-E card running on CUDA 12.0. Batch normalization is implemented after each convolutional layer to unify the scale of each Graphical Processing Unit's (GPU's) training data.
\begin{table}[!t]
\centering
\caption{{Specifications for the CNN used to perform SEI.}} \label{tab:sei_arch}
\begin{subfigure}[{CNN specifications for time- and frequency-based SEI.}]{\label{tab:sei_time_freq_arch}
\includegraphics[width=\columnwidth]{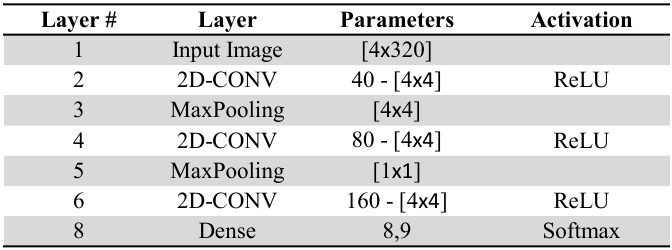}}
\end{subfigure}
\medskip
\begin{subfigure}[{CNN specifications for Gabor-based SEI.}]{\label{tab:sei_gabor_arch}
\includegraphics[width=\columnwidth]{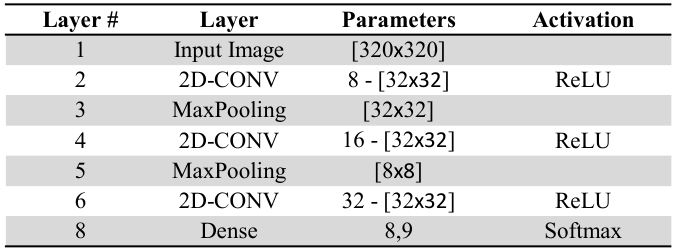}}
\end{subfigure}
\vspace{-2mm}
\end{table}
\subsection{The ``Coffee Shop'' Scenario%
\label{sec:coffee_shop_methods}}
\begin{figure}[!t]
\centering
\includegraphics[width=0.95\columnwidth]{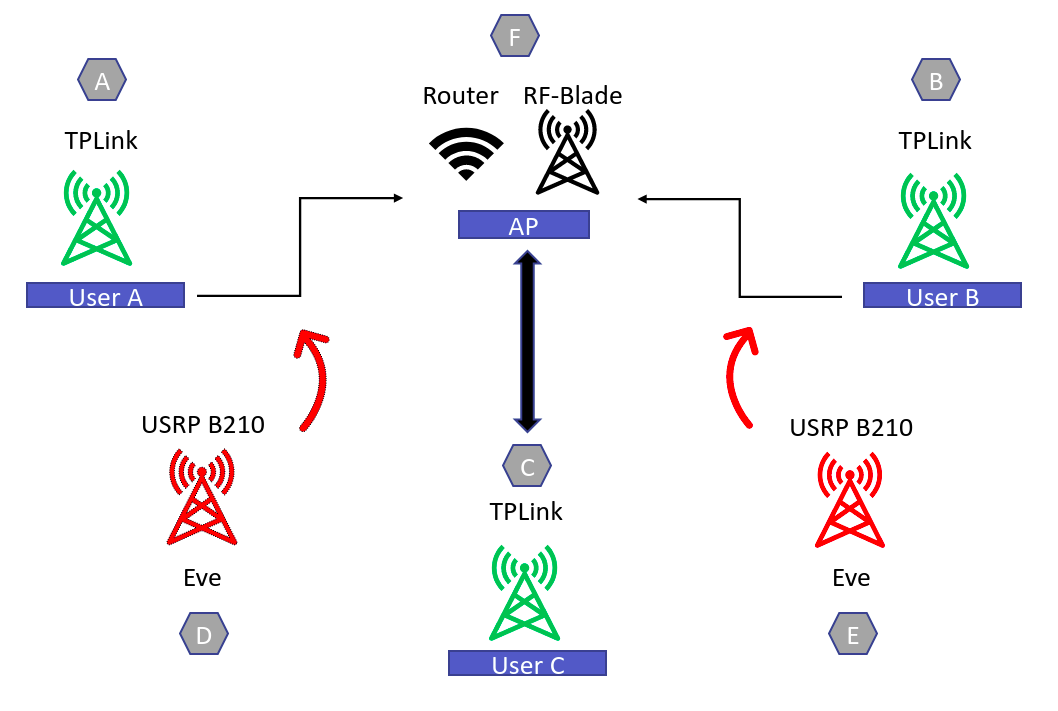}
\caption{The coffee shop scenario setup in our laboratory for in-situ testing of Eve's GAN-based SEI mimicry countermeasure. The red arrows indicate the user emitter whose SEI features are being mimicked by Eve from its indicated location.}
\label{fig:coffee_shop_setup}
\end{figure}

Figure~\ref{fig:coffee_shop_setup} shows the setup used for this scenario, which consists of an IEEE 802.11a Wi-Fi network comprised of three authorized emitters--designated User A, User B, and User C--and an Access Point (AP). Each user is a TP-Link AC1300 USB Wi-Fi adapter and a computer running Ubuntu Linux 16.0. The AP consists of an ASUS RT-AX56U IEEE 802.11a Wi-Fi router and Nuand BladeRF SDR powered by a Raspberry Pi 4 model B. The AP is located roughly equidistant from each of the three authorized emitters. The Nuand BladeRF SDR collects signals transmitted by each emitter and performs SEI using a CNN and the time representation of Equation~\eqref{eqn:preamble_time_rep} to authenticate a given user's identity. An SFTP server provides a file transfer service for the users in the coffee-shop network and is co-located with User C. User A and User B are configured to transmit binary files to User C and User C transmits binary files to User A. The AP forwards Wi-Fi frames from User A and User B to the SFTP server while the BladeRF SDR collects the same frames to train the CNN for SEI.\\ 
\indent Eve uses a B210 or two HackRFs controlled by an NVIDIA Jetson Nano Developer Kit. Regardless of the SDR, Eve (i) repeatedly transmits the same frame from Location~{D} or E, (ii) receives Wi-Fi frames transmitted by itself and the three users, (iii) processes all Wi-Fi frames as described in Section~\ref{sec:signal_collect}, (iv) trains the GAN using backpropagation on an NVIDIA Tesla K40m GPU and quantizes the model using 12~{bit} resolution, and (v) uses the trained $G$ to attack the SEI process at Location~{F}. Eve mimics User~{A's} and User~{C's} SEI features from Location~{D} while User~{B's} SEI features are mimicked when Eve is at Location~{E}. During its attack, Eve removes the last two rows of the $G$'s output tensor because they are the IQ samples' magnitude and phase and are no longer needed. Ten thousand preambles are collected for User~{A}, User~{B}, User~{C}, and Eve when its SEI countermeasure is turned off. Eight thousand preambles are randomly selected for training and validation while the rest are used for testing. The collected preambles have an estimated SNR of between 10~{dB} and 12~{dB}.
\begin{table}[!b]
\centering
\vspace{-3mm}
\caption{Identity Authentication Outcomes.}
\label{tab:VerifyErrors}
\includegraphics[width=\columnwidth]{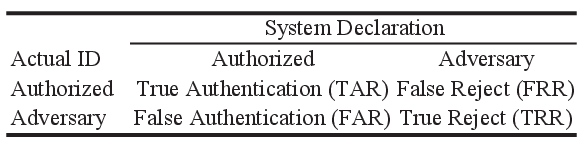}
\end{table}
\begin{table*}[!t]
\caption{FAR and TRR--presented in each cell as ``FAR/TRR'' using percentages--when each SEI process is and is not trained using a decoy emitter's RF-DNA fingerprints at SNR values of $30$~{dB} and $9$~{dB}. SEI processes that result in a FAR above $12.5${\%} (a.k.a., a guess) and a TRR above $90${\%} are marked using \textcolor{red}{\textit{italicize-red}} and \textcolor{mygreen}{\textbf{bold-green}} font, respectively. ``RE'', ``AE'', and ''GAN'' indicate Eve's replay, AE, and GAN-based mimicry countermeasures, respectively.}
\label{tab:mimic_results}
\begin{minipage}[b]{\textwidth}
\centering
\subfigure[SEI using preambles \textit{without} the decoy present at SNR$=$30~{dB}.]{\label{fig:sec_rej_1}\includegraphics[width=0.49\textwidth]{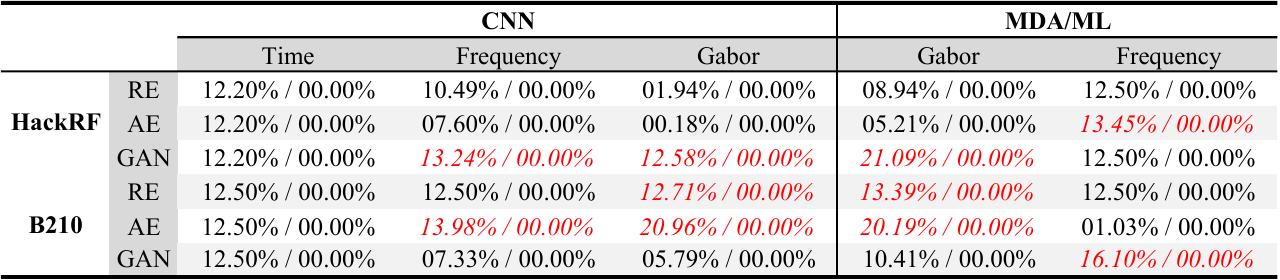}}
\subfigure[SEI using preambles \textit{with} the decoy present at SNR$=$30~{dB}.] {\label{fig:sec_rej_3}\includegraphics[width=0.49\textwidth]{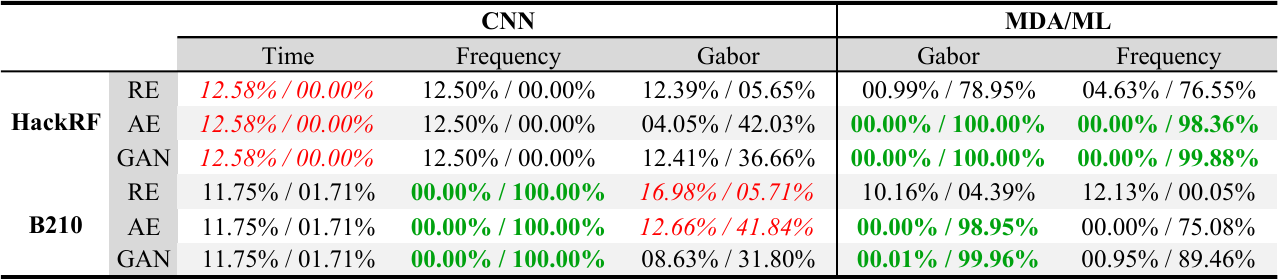}}
\subfigure[SEI using preambles \textit{without} the decoy present at SNR$=$9~{dB}.] {\label{fig:sec_rej_2}\includegraphics[width=0.49\textwidth]{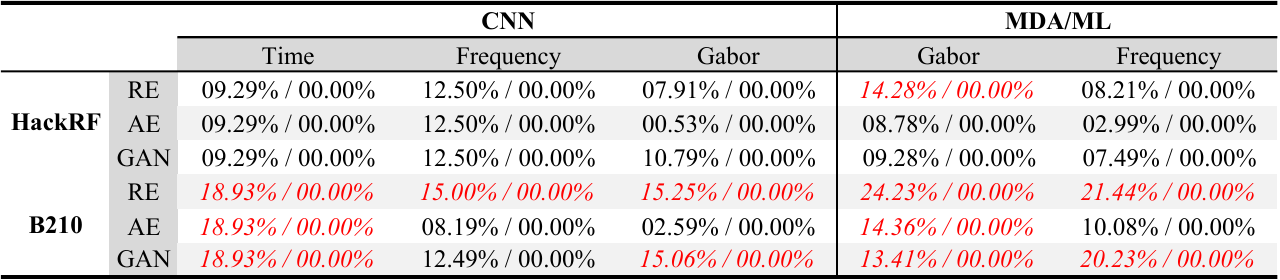}}
\subfigure[SEI using preambles \textit{with} the decoy present at SNR$=$9~{dB}.] {\label{fig:sec_rej_4}\includegraphics[width=0.49\textwidth]{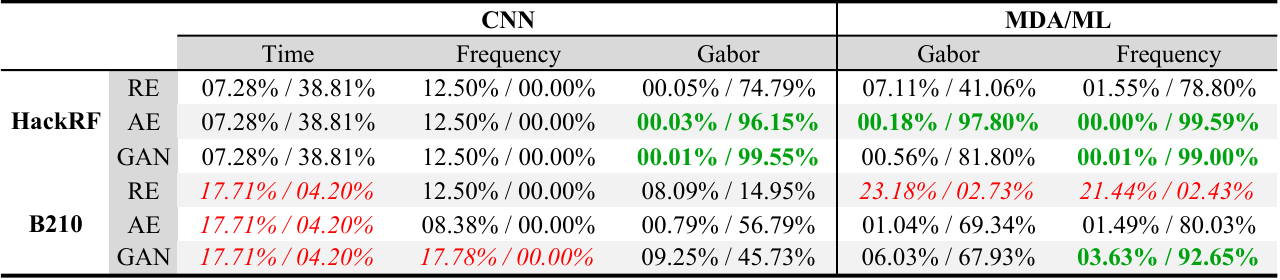}}
\end{minipage}
\begin{minipage}[b]{\textwidth}
\centering
\subfigure[SEI using residual preambles \textit{without} the decoy present at SNR$=$30~{dB}.]{\label{fig:sec_rej_5}\includegraphics[width=0.49\textwidth]{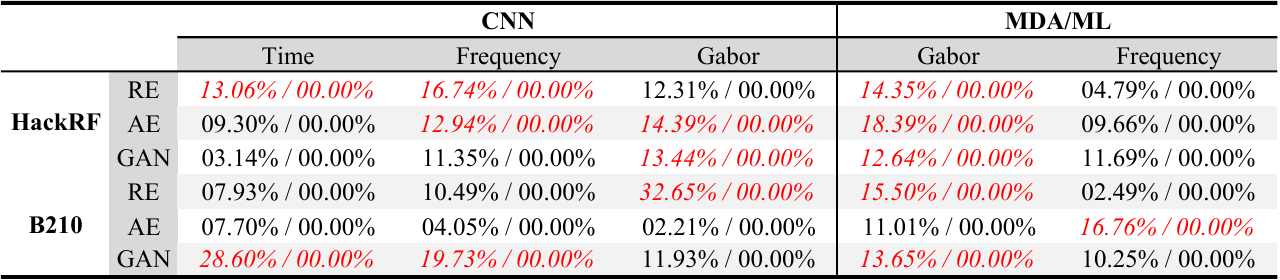}}
\subfigure[SEI using residual preambles \textit{with} the decoy present at SNR$=$30~{dB}.] {\label{fig:sec_rej_7}\includegraphics[width=0.49\textwidth]{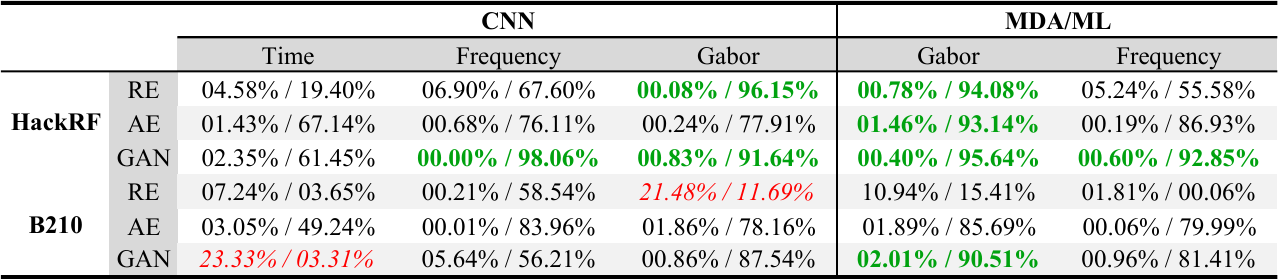}}
\subfigure[SEI using residual preambles \textit{without} the decoy present at SNR$=$9~{dB}.] {\label{fig:sec_rej_6}\includegraphics[width=0.49\textwidth]{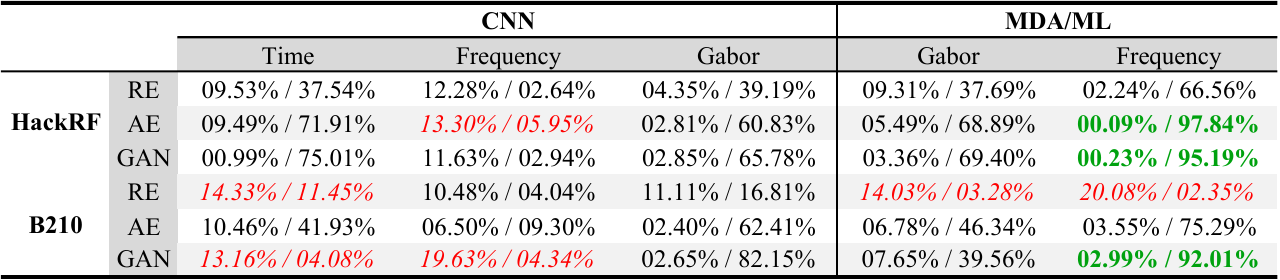}}
\subfigure[SEI using residual preambles \textit{with} the decoy present at SNR$=$9~{dB}.] {\label{fig:sec_rej_8}\includegraphics[width=0.49\textwidth]{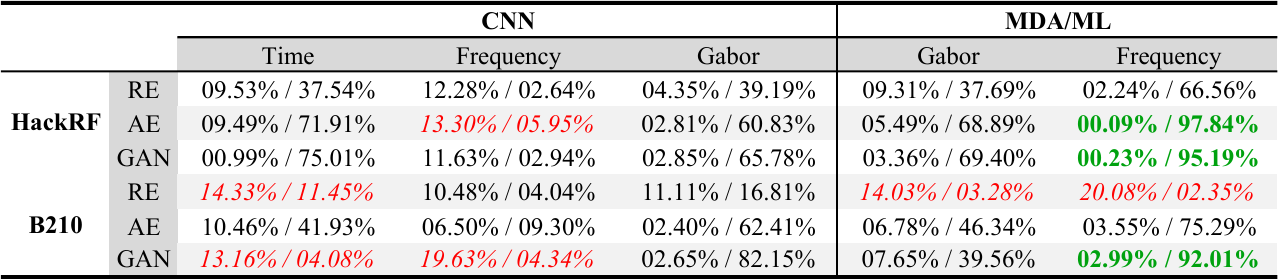}}
\end{minipage}
\vspace{-5mm}
\end{table*}
\section{RESULTS%
\label{sec:results}}
Eve's SEI countermeasure is evaluated using the outcomes in Table~\ref{tab:VerifyErrors}. Of particular interest are the False Authentication Rate (FAR) and True Reject Rate (TRR), which indicate the rate at which Bob incorrectly authenticates Eve as an authorized emitter and correctly identifies Eve as an unauthorized emitter, respectively. FAR is the number of times Eve is incorrectly authenticated as one of the authorized emitters divided by the total number of transmissions made by Eve. The TRR is the number of times Eve is correctly identified as an unauthorized emitter divided by the number of transmissions made by Eve. Since the SEI processes perform classification, the TRR will be 0\% whenever a decoy emitter's RF-DNA fingerprints \textit{are not} part of the training set (i.e., Eve is unknown to the SEI process prior to its attack), because the decoy represents Eve and serves as the unauthorized emitter. {This is a byproduct of the classification process, which assigns every input RF-DNA fingerprint to one of the known classes no matter how poorly the RF-DNA fingerprint matches the class to which it is assigned.} This classification drawback is addressed by either including decoy emitter RF-DNA fingerprints (see Table~\ref{tab:mimic_results}) or Eve's unmodified signals (see Section~\ref{sec:results_coffee_shop}) within the SEI processes' training sets.\\
\indent Table~\ref{tab:mimic_results} shows FAR and TRR--presented in each cell as ``FAR/TRR'' using percentages--when each SEI process is and is not trained using a decoy emitter's RF-DNA fingerprints at an SNR of $30$~{dB} or $9$~{dB}. MDA/ML-based SEI uses either Gabor-based RF-DNA fingerprints (see Section~\ref{sec:hand_rf_dna}) or the frequency representation of Equation~\eqref{eqn:preamble_freq_rep}. The CNN-based SEI learns RF-DNA fingerprints from the time, frequency, or Gabor-based images described in Section~\ref{sec:dl_rf_finger}. At an SNR$=$30~{dB}, Table~\ref{fig:sec_rej_1} shows that a HackRF-based Eve employing replay and AE-based mimicry are classified at a FAR$=$$12.5${\%} by the CNN-based SEI using all three preamble representations. When the HackRF-based Eve uses GAN-based mimicry, the FAR increases to $13.24${\%} and $12.58${\%} for the frequency and Gabor image representations, respectively. The FAR$=$$13.45${\%} for MDA/ML-based SEI uses the frequency representation to classify the HackRF-based Eve employing AE-based mimicry. When Eve uses the HackRF and GAN-based mimicry, MDA/ML-based SEI classifies Eve at a FAR$=$$21.09${\%} when Gabor-based RF-DNA fingerprints. A B210-based Eve employing replay mimicry results in a FAR$=$$12.71${\%} against CNN-based SEI using Gabor images. When the B210 employs AE-based mimicry, it has a FAR of $13.98${\%} and $20.96${\%} against the CNN using frequency and Gabor coefficients, respectively, and a FAR of $20.19${\%} against the MDA/ML classifier using handcrafted fingerprints. When using GAN-based mimicry, the B210 only has a FAR above the guess threshold when being classified by the MDA/ML using frequency coefficients with a rate of $16.10${\%}.\\
\indent Table~\ref{fig:sec_rej_3} shows FAR and TRR for preambles collected at $30$~{dB} and SEI processes trained \textit{with} decoy emitter preambles. A HackRF-based Eve using replay, AE, or GAN mimicry achieves a FAR$=$$12.58${\%} against the CNN and time-based RF-DNA fingerprinting SEI process. A HackRF-based Eve using AE-based mimicry is defeated at a TRR of $100$\% and $98.36$\% when MDA/ML-based SEI uses Gabor and frequency-based RF-DNA fingerprints. When Eve switches from the AE-based mimicry to GAN-based, MDA/ML-based SEI achieves a TRR of $100$\% and $99.88$\% when using Gabor and frequency-based RF-DNA fingerprints. When Eve uses a B210 and any of the mimicry methods, CNN-based SEI rejects all of them at a TRR$=$100\% when using frequency-based RF-DNA fingerprints. However, the Gabor image and CNN-based SEI process does not fair as well with a FAR of $16.98${\%} and $12.66${\%} against a B210-based Eve employing replay and AE-based mimicry. When Eve uses a B210 and either AE or GAN-based mimicry, MDA/ML-based SEI rejects Eve at TRRs of $98.95${\%} and $99.96${\%} using Gabor-based RF-DNA fingerprints.
\\
\indent Table~\ref{fig:sec_rej_2} shows FAR and TRR for preambles collected at $9$~{dB} and SEI processes trained \textit{without} decoy emitter preambles. HackRF-based Eve uses replay mimicry to achieve a FAR$=$$14.28${\%} against MDA/ML-based SEI that uses Gabor-based RF-DNA fingerprints. For all other cases, a HackRF-based Eve does not achieve a FAR$>$$12.5${\%}. A B210-based Eve--using a replay countermeasure against CNN-based SEI--achieves a FAR of $18.93${\%}, $15.00${\%}, and $15.25${\%}  against time, frequency, and Gabor image RF-DNA fingerprinting, respectively. This same Eve configuration achieves a FAR of $24.23${\%} and $21.44${\%} against MDA/ML-based SEI that uses Gabor and frequency representation RF-DNA fingerprints, respectively. A FAR$=$$18.93${\%} is achieved by the B210-based Eve using AE-based mimicry against CNN-based SEI using the time representation of Equation~\eqref{eqn:preamble_time_rep}. When employing AE-based mimicry against an MDA/ML and Gabor-based SEI process, a B210-based Eve achieves a FAR$=$$14.36${\%}. The B210-based Eve achieves a FAR of $18.93${\%} and $15.06${\%} using GAN-based mimicry against CNN SEI processes using time and Gabor image RF-DNA fingerprints as well as FAR values of $13.41${\%} and $20.23${\%} against the MDA/ML built SEI processes using Gabor and frequency-based RF-DNA fingerprints.
\\
\indent Table~\ref{fig:sec_rej_4} shows FAR and TRR for preambles collected at $9$~{dB} and SEI processes trained \textit{with} decoy emitter preambles. A HackRF-based Eve using AE or GAN-based mimicry is successfully rejected at a TRR of $96.15${\%} or $99.55${\%} by CNN-based SEI using Gabor images. The MDA/ML SEI process also rejects a HackRF-based Eve--using AE mimicry--at a TRR of $97.8$\% and $99.59$\% when using Gabor and frequency-based RF-DNA fingerprints. When using frequency-based RF-DNA fingerprints the MDA/ML SEI process rejects a HackRF-based Eve using GAN mimicry at a TRR$=$$99$\%. When Eve employs a B210 achieves a FAR$=$$17.71$\% using all three mimicry countermeasures against the CNN SEI process that extracts RF-DNA fingerprints from the time representation. The FAR$=$$17.78${\%} when the B210-based Eve uses GAN-based mimicry against the SEI process using CNN and the frequency representation. A FAR of $23.18${\%} and $21.44${\%} results when the B210-based Eve uses replay mimicry against the SEI process using MDA/ML and Gabor-based RF-DNA fingerprints. The SEI process that uses MDA/ML and the frequency representation rejects the B210-based Eve using GAN-based mimicry at a TRR$=$92.65\%.
\\
\indent Table~\ref{fig:sec_rej_5} shows FAR and TRR for residual preambles collected at $30$~{dB} and SEI processes trained \textit{without} decoy emitter residual preambles. The residual preamble is generated by subtracting an ideal preamble from each received preamble as detailed in~\cite{tyler2022assessing}. When using replay mimicry, a HackRF-based Eve achieves a FAR of $13.06${\%} and $16.74${\%} against CNN-based SEI extracting RF-DNA fingerprints from the time or frequency representation. Employing AE and GAN-based mimicry the HackRF-based Eve achieves FARs of $14.39${\%} and $13.44${\%} against CNN-based SEI that extracts RF-DNA fingerprints from the Gabor images. The HackRF-based Eve also achieves FARs of $14.35${\%}, $18.39${\%}, and $12.64${\%} when employing replay, AE, and GAN-based mimicry against MDA/ML-based SEI. The B210-based Eve achieves a FAR of $32.65${\%} when using replay mimicry against the CNN and Gabor image-based SEI process. Using GAN-based mimicry, the B210-based Eve achieves FARs of $28.60${\%} and $19.73${\%} against CNN-based SEI that learns RF-DNA fingerprints from the time or frequency representation. Replay and GAN-based mimicry achieve FARs of $15.5${\%} and $13.65${\%} against the MDA/ML-SEI process that uses Gabor-based RF-DNA fingerprints and Eve uses a B210. B210-based Eve uses AE-based mimicry to achieve a FAR$=$$16.76${\%} versus MDA/ML-based SEI that uses the preambles' frequency representation.
\\
\indent Table~\ref{fig:sec_rej_7} shows FAR and TRR for residual preambles collected at $30$~{dB} and SEI processes trained \textit{with} decoy emitter residual preambles. Using replay mimicry and a HackRF, Eve is rejected at a TRR$=$$96.15${\%} by CNN SEI based on Gabor images. CNN SEI rejects GAN-based mimicry by HackRF-based Eve at TRRs of $98.06$\% and $91.64$\% using the frequency and Gabor image representations, respectively. HackRF-based Eve is rejected at TRRs of $94.08${\%}, $93.14${\%}, and $95.64${\%} by the Gabor-based, MDA/ML SEI process when Eve uses replay, AE, and GAN-based mimicry. HackRF-based Eve is rejected at a TRR$=$$92.85${\%} when using GAN-based mimicry against MDA/ML SEI that uses the preambles' frequency representations. When the B210-based Eve uses replay mimicry against CNN-based SEI using Gabor images Eve is achieves a FAR$=$$21.48${\%}. The B210-based Eve is rejected at a TRR$=$$90.51$\% when using GAN-based mimicry against MDA/ML SEI that uses Gabor-based RF-DNA fingerprints.
\\
\indent Table~\ref{fig:sec_rej_6} shows FAR and TRR for residual preambles collected at $9$~{dB} and SEI processes trained \textit{without} decoy emitter residual preambles. A HackRF-based Eve using replay mimicry achieves FARs of $14.08${\%}, $18.68${\%}, and $14.44${\%} against the CNN-based SEI processes using time, frequency, and Gabor image representations. HackRF-based Eve's AE-based mimicry achieves FARs of $16.63${\%} and $20.55${\%} when CNN-based SEI uses time or frequency representations. When HackRF-based Eve uses GAN-based mimicry a FAR$=$$12.78${\%} results against CNN-based SEI using the frequency representation. HackRF-based Eve using replay and AE-based mimicry achieves FARs of $14.21${\%} and $14.43${\%} against MDA/ML SEI that uses Gabor-based RF-DNA fingerprints. For the replay, AE, and GAN-based mimicry, the B210-based Eve results in FARs of $13.46${\%}, $19.46${\%}, and $14.53${\%} versus the CNN SEI process that uses the time representation. When Eve switches to a B210 a FAR of $16.39${\%} is achieved for GAN-based mimicry versus the frequency representation-reliant CNN-based SEI process. B210-based Eve achieves FARs of $14.36${\%} and $18.90${\%} using replay and GAN-based mimicry against CNN-based SEI that uses Gabor images. FARs of $14.36${\%} and $18.90${\%} are achieved by the B210-based Eve using replay mimicry against Gabor and frequency representation-reliant MDA/ML-based SEI processes. The FAR increases to $19.54${\%} for the B210-based Eve using GAN-based mimicry against the frequency representation using MDA/ML SEI process.
\\
\indent Table~\ref{fig:sec_rej_8} shows FAR and TRR for residual preambles collected at $9$~{dB} and SEI processes trained \textit{with} decoy emitter residual preambles. HackRF-based Eve achieves a FAR$=$$13.30${\%} using AE-based mimicry against CNN SEI that uses the frequency representation. When HackRF-based Eve is rejected at a TRR$=$$97.84${\%} and $95.19${\%} when using AE and GAN-based mimicry against MDA/ML SEI that uses the frequency representation. FARs of $14.33${\%} and $13.16${\%} result for CNN SEI that uses the time representation to classify the replay and GAN-based mimicked preambles of the B210-based Eve. The B210-based Eve achieves a FAR$=$$19.63${\%} when using GAN-based mimicry against the frequency representation-reliant CNN-based SEI process. B210-based Eve's replay mimicry achieves FARs of $14.03${\%} and $20.08${\%} when employing replay mimicry against MDA/ML SEI that uses Gabor and frequency representation-based RF-DNA fingerprints, respectively. B210-based Eve is rejected at a TRR$=$$92.01${\%} when using GAN-based mimicry against MDA/ML SEI built on the preambles' frequency representations.

{Overall, Eve's SEI mimicry countermeasure is most effective at 9~{dB} and when the SEI process is trained using the authorized emitters' residual preambles (i.e., decoy emitter residual preambles are not utilized in SEI process training). This is not surprising in that most SEI processes struggle to separate even known emitters as SNR degrades. So, the introduction of SEI mimicked features and no point of reference (i.e., no decoy emitter present) only exacerbates the issue. When not using decoy emitter preambles, the SEI is most effective in defeating Eve's SEI mimicry countermeasure when using the preambles' or residual preambles' frequency representations and the MDA/ML classifier. Lastly, SWaP-C does negatively impact the effectiveness of Eve's SEI mimicry countermeasures. Use of the lower SWaP-C HackRF One SDR leads to degraded SEI mimicry countermeasure effectiveness.}
\subsection{Results: ``Coffee Shop Scenario''%
\label{sec:results_coffee_shop}}
In addition to SDR SWaP-C, the ``coffee shop scenario'' considers two cases, thus the results for this scenario are presented based on these two cases. Regardless of the case, the success of Eve’s SEI countermeasure is determined by calculating the ratio between the number of times the modified preambles--output by Eve's Location D trained $G$—-are classified as originating from User A and the total number of SEI process decisions. Eve’s success rate is calculated the same way when Eve operates at Location D but mimics the SEI features of User C or Location E and mimics User B.
\subsubsection{Coffee Shop Scenario Case~{\#1}%
\label{sec:coffee_results_case1}}

\begin{table}[!b]
\centering
\caption{\textit{Coffee Shop Scenario Case~{\#1}:} Percent classification for Eve using SEI mimicry and its unmodified signals \underline{\textbf{are}} used to train the SEI process. ``GAN'' and ``No GAN'' indicate SEI mimicry is ``on'' and ``off'', respectively.} \label{tab:coffee_shop_case1_results}
\begin{subfigure}[Eve using a HackRF One SDR.]{\label{tab:coffee_shop_case1_hack_results}
\includegraphics[width=\columnwidth]{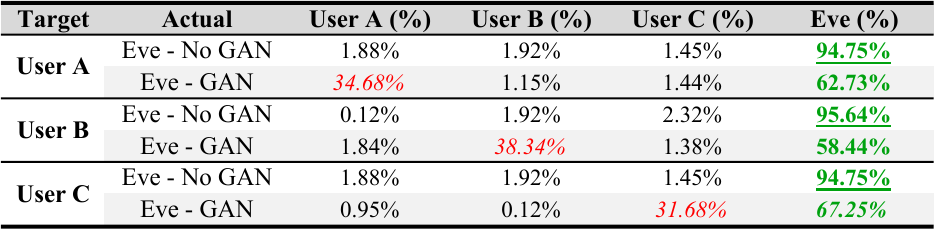}}
\end{subfigure}
\medskip
\begin{subfigure}[Eve using a B210 SDR.]{\label{tab:coffee_shop_case1_b210_results}
\includegraphics[width=\columnwidth]{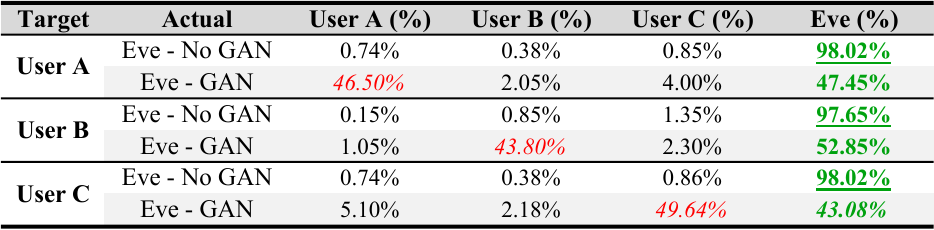}}
\end{subfigure}
\end{table}

The SEI process \textit{is} trained using Eve’s unmodified signals, thus it can identify Eve when Eve's SEI countermeasure is {\textit{off}. The AP saves Eve's preambles transmitted from Location~{D} and Location~{E} using separate files. Case~{\#1} results are presented in Table~\ref{tab:coffee_shop_case1_hack_results} and Table~\ref{tab:coffee_shop_case1_b210_results} for a HackRF and B210-based Eve, respectively.\\
\indent Table~\ref{tab:coffee_shop_case1_b210_results} shows the FAR is 46.5\% and 49.64\% when Eve launches the SEI countermeasure--from Loaction~{E}--using a B210 to mimic User~{A's} and User~{C's} SEI features, respectively. From Location~{D}, Eve’s SEI countermeasure degrades the SEI process’ ability to distinguish Eve from User~{A} and User~{C} by almost 50\%. When Eve mimics the SEI features of User~{B} from Location~{E}, the SEI process identifies Eve correctly 43.8\% of the time, which is in stark contrast to the 97.65\% correct identification rate when Eve is not mimicking User~{B} (a.k.a., “Eve - No GAN”). The SEI process still performs better than a guess, but its ability to classify Eve is reduced. Eve’s attack is less successful when operating from Location~{E}, which is attributed to the SEI process' training set not containing any of Eve’s unmodified signals (a.k.a., SEI mimicry is \textit{off}) transmitted from Location~{E}. We suspect the success rate would change if this were not the case. \\
\begin{table}[!b]
\centering
\caption{\textit{Coffee Shop Scenario Case~{\#2}:} Percent correct classification for Eve using SEI mimicry and its unmodified signals \underline{\textbf{are not}} used to train the SEI process. ``GAN'' and ``No GAN'' indicate SEI mimicry is ``on'' and ``off'', respectively.} \label{tab:coffee_shop_case2_results}
\begin{subfigure}[Eve using a HackRF One SDR.]{\label{tab:coffee_shop_case2_hack_results}
\includegraphics[width=\columnwidth]{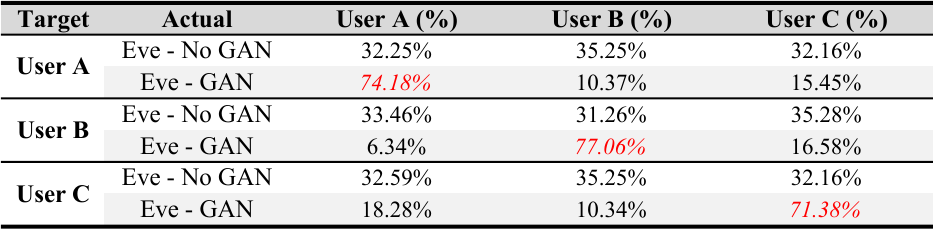}}
\end{subfigure}
\medskip
\begin{subfigure}[Eve using a B210 SDR.]{\label{tab:coffee_shop_case2_b210_results}
\includegraphics[width=\columnwidth]{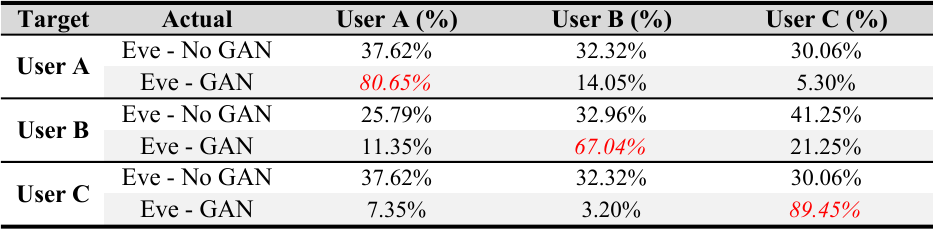}}
\end{subfigure}
\end{table}
\begin{table*}[!t]
\centering
\caption{FAR and TRR--presented in each cell as ``FAR/TRR'' using percentages--when a DAE precedes the SEI process' trained using a decoy emitter's RF-DNA fingerprints at an SNR $9$~{dB}. SEI processes that result in a FAR above $12.5${\%} (a.k.a., a guess) and a TRR above $90${\%} are marked using \textcolor{red}{\textit{italicize-red}} and \textcolor{mygreen}{\textbf{bold-green}} font, respectively.  ``RE'', ``AE'', and ''GAN'' indicate Eve's replay, AE, and GAN-based mimicry countermeasures, respectively.}
\includegraphics[width=0.8\textwidth]
{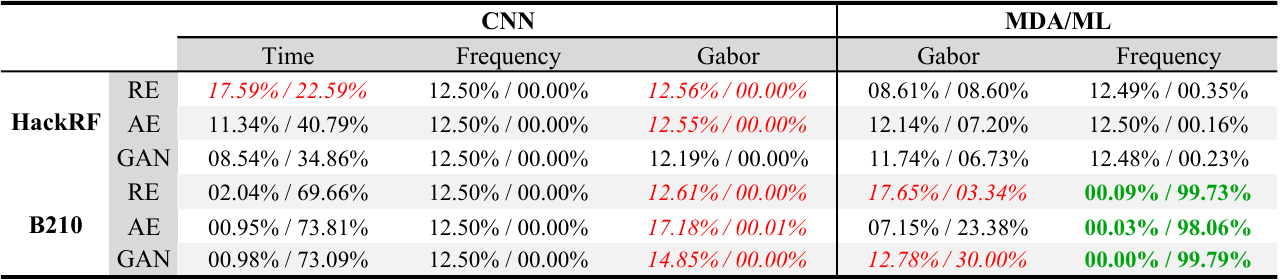}
\label{tab:sei_9dB_ae}
\end{table*}
\indent When Eve uses a HackRF SDR (see Table~\ref{tab:coffee_shop_case1_hack_results}), the SEI process accurately classifies Eve's signals at a percent correct classification rate of 94.75\%, 94.75\%, and 95.64\% when Eve \textit{is not} mimicking (a.k.a., ``Eve - No GAN'') the SEI features of User A, User B, and User C, respectively. Thus, the SEI process easily distinguishes Eve from the authorized users. However, once Eve employs an SEI countermeasure (a.k.a., ``Eve - GAN''), then the SEI process' accuracy is degraded. When mimicking User A's SEI features, the SEI process' ability to distinguish Eve from User A is reduced by 34.68\%, thus Eve is correctly identified 62.73\% of the time. When mimicking User C, Eve's GAN-based mimicry countermeasure reduces the SEI process accuracy from 94.75\% to 67.25\%, thus Eve's signals are being misclassified as originating from User C 31.68\% of the time. Eve's mimicry of User C is the least effective while Eve's SEI countermeasure is most effective when Eve is mimicking User B's SEI features. When Eve mimics User B, SEI process accuracy is reduced from 95.64\% to 58.44\%. Overall, Eve's SEI countermeasure hinders the SEI process' ability to distinguish Eve from the targeted emitter, however not to the extent that they are indistinguishable. We also note that the results in Table~\ref{tab:coffee_shop_case1_hack_results} are poorer than those in Table~\ref{tab:coffee_shop_case1_b210_results} that correspond to a USRP B210-based Eve. The reduced performance--exhibited in Table~\ref{tab:coffee_shop_case1_hack_results}--is attributed to the SWaP-C constraints imposed on Eve when using a HackRF in lieu of a B210. The most prominent constraint is a lower sampling rate of 20~{MHz} versus the B210's 40~{MHz}. The sampling rate had to be lowered to keep the HackRF from ceasing transmissions once it was connected to the NVIDIA Jetson Nano. If the sampling rate is not reduced, then a majority of the collected Wi-Fi frames' samples are missing, which negatively impacts preamble detection and extraction.

\subsubsection{Coffee Shop Scenario Case~{\#2}%
\label{sec:coffee_results_case2}} 
The SEI process \textit{is not} trained using Eve’s unmodified signals. The SEI process CNN is trained using only authorized emitter signals and Eve is an unknown prior to launching a mimicry attack. Table~\ref{tab:coffee_shop_case2_results} shows Eve is classified as each user about one-third of the time. The CNN is guessing which authorized user transmitted the signals, thus there is no bias toward one user over another that would bias the results when Eve mimics each user's SEI features.\\
\indent When Eve mimics User~{A's} SEI features, the SEI process classifies them as originating from User~{A} at a 74.18\% accuracy, Table~\ref{tab:coffee_shop_case2_hack_results} The remaining signals are classified as originating from User~{B} and User~{C} at accuracies of 10.37\& and 15.45\%, respectively. When mimicking User~{B}, Eve's SEI countermeasure leads its modified signals to be classified as originating from User~{B} at an accuracy of 77.08\%. This represents Eve's highest accuracy when using a HackRF to implement SEI mimicry and is 9.68\% higher than Eve's accuracy when using a B210, Table~\ref{tab:coffee_shop_case2_b210_results}. The HackRF-based Eve's lowest accuracy of 71.38\% occurs when mimicking User~{C}, Table~\ref{tab:coffee_shop_case2_hack_results}. However, the B210-based Eve results in a lower classification accuracy of 67.4\% when mimicking User~{B's} SEI features, Table~\ref{tab:coffee_shop_case2_b210_results}. This performance difference may be due to the HackRF's lower sampling rate. A lower sampling rate means some SEI nuances are missing from the emitters' sampled signals, thus impeding the SEI process' ability to discriminate between emitters. This is of even greater importance when the SEI process is trained without Eve's unmodified signals because the CNN needs as much information as possible to achieve reliable discrimination.\\
\indent Table~\ref{tab:coffee_shop_case2_b210_results} shows the SEI countermeasure success rate for the B210-based Eve. When Eve transmits with the SEI countermeasure \textit{turned off} (a.k.a., ``Eve - No GAN''), the SEI process is roughly guessing as to the originator of the signals. However, once Eve uses the SEI countermeasure (a.k.a., ``Eve - GAN''), the SEI process mostly assigns Eve the identity of the user being mimicked. The SEI countermeasure’s greatest success is when User~{C} is mimicked, which results in an 89.45\% identification rate versus only 30.1\% when the countermeasure is \textit{turned off}. The SEI countermeasure’s lowest success rate of 67.4\% occurs when Eve mimics User~{B's} SEI features. This still represents a roughly 44\% increase in identifying Eve as User~{B}. This is interesting when considering that when Eve is at Location~{E} and its SEI countermeasure is \textit{turned off} because the SEI process identifies Eve as User~{B} roughly 33\% of the time. This is not the highest rate—-the highest rate is 41.25\% which corresponds to User~{C}-—which shows the SEI countermeasure alters Eve’s SEI features sufficiently to change which user Eve ``looks most like''. Eve goes from ``looking most like'' User~{C} to ``looking most like'' User~{B} when the SEI countermeasure goes from turned \textit{off} to \textit{on}.
\subsection{Results: Denoised Signals with MDA/ML-based SEI%
\label{sec:denoise_results}}
Motivated by our work in~\cite{tyler2021simplified}, a DAE is added to the SEI process to improve rejection of Eve's SEI mimicry attacks an SNR$=$$9$~{dB}. The DAE is trained using $30$~{dB} and $9$~{dB} collected preambles with the $30$~{dB} preambles used as the targeted output. The $9$~{dB} preambles are matched to the same emitter's $30$~{dB} preambles. At an SNR$=$$9$~{dB}, Eve's mimicry attacks are detected at the highest rate when a ``decoy'' emitter's RF-DNA fingerprints are learned by the SEI process, Table~\ref{tab:sei_9dB_ae}. When employing replay mimicry, a HackRF-based Eve achieves FARs of 17.59{\%} and 12.56{\%} versus a CNN-based SEI process that uses time and Gabor-based RF-DNA fingerprints, respectively. A FAR$=$$12.55${\%} results when Eve employs AE-based mimicry and a HackRF against CNN-based SEI using Gabor-based RF-DNA fingerprints. However, CNN-based SEI using Gabor-based RF-DNA fingerprints corresponds to FARs of $12.61${\%}, $17.18${\%} and $14.85${\%} when Eve uses a B210 and replay, AE, and GAN-based mimicry, respectively. When Eve uses a B210 with replay or GAN-based mimicry against MDA/ML-based SEI that uses handcrafted RF-DNA fingerprints FARs of $17.65${\%} and $12.78${\%} result, respectively. However, MDA/ML-based SEI achieves TRRs of $99.73${\%}, $98.06${\%}, and $99.79${\%} against a B210-based Eve using replay, AE, or GAN-based SEI countermeasures, respectively.
\section{CONCLUSION%
\label{sec:conlusion}}
This work considers a strong adversary (a.k.a., Eve) that uses ``off-the-shelf'' software tools, DL, and SDR to perform digital and SEI mimicry to deceive an SEI-based monitor (a.k.a., Bob) to gain network access. SEI processes employing classification without knowledge of Eve's organic features or those of a representative (a.k.a., a decoy) incorrectly authenticate Eve as an authorized user. The success of Eve's SEI mimicry is higher when using a higher SWaP-C SDR (e.g., a B210 versus HackRF One). FAR is higher at $9$~{dB} than at $30$~{dB}, thus it is advantageous for Eve to launch SEI mimicry at lower SNRs to increase SEI confusion. Eve can intentionally add noise to its signals prior to transmission or lower its transmit power to change its signals' SNR. The effectiveness of Eve's SEI mimicry is degraded when the SEI process is trained using decoy emitter signals. TRR is lower for all SEI processes, both SDRs and the three SEI mimicry types at $9$~{dB}. Denoising signals prior to SEI does not decrease the success of Eve's SEI mimicry countermeasures. This may be due to the DAE ``washing out'' essential emitter-specific features used by the SEI process to distinguish between emitters. Overall, handcrafted SEI--without DAE--using Gabor-based RF-DNA fingerprints provides the best FAR/TRR performance albeit at a greater computational cost. Many of these observations are validated by assessing Eve's SEI mimicry countermeasure in a ``coffee shop'' setting. Future work will consider open-set SEI to remove the need for decoy emitter signals as well as one-to-one identity verification to overcome classification's ``forced decision'' drawback. 
%

\bibliographystyle{IEEEtran}
\bibliography{002_OJCOMS_Refs__v01.bib}

\end{document}